\documentclass[a4paper,12pt,openright, singlespace]{article}
\usepackage{graphics,subcaption,graphicx}
\usepackage{algorithm,algpseudocode}
\usepackage{amsmath,amsfonts,amssymb,amsthm}
\usepackage{epstopdf}
\usepackage{color}
\usepackage{comment} 

\ifpdf
  \DeclareGraphicsExtensions{.eps,.pdf,.png,.jpg}
\else
  \DeclareGraphicsExtensions{.eps}
\fi
\newcommand{\D}{\displaystyle}

\newcommand{\dsum}{\D\sum}
\newcommand{\dint}{\D\int}

\newcommand{\uk}{\textbf{u}}

\newcommand{\vk}{\textbf{v}}

\newcommand{\zk}{\textbf{z}}

\renewcommand{\Re}{\text{Re}}
\newcommand{\fk}{\textbf{f}}

\newcommand{\intO}[1]{\dint_\Omega #1\; d\Omega}

\newcommand{\cD}{\mathcal{D}}
\newcommand{\cN}{\mathcal{N}}
\newcommand{\R}{\mathbb{R}}

\newcommand{\normld}[1]{\|#1\|_{0,2,\Omega}}

\newcommand{\Th}{\mathcal{T}_h}

\begin{document}

\title{A Reduced Basis LES turbulence model based upon Kolmogorov’s equilibrium turbulence theory}
\author{
Alejandro Bandera Moreno\thanks{Departamento de Ecuaciones Diferenciales y Análisis Numérico, 
and Instituto de Matemáticas de la Universidad de Sevilla (IMUS)
, Apdo. de correos 1160, Universidad de Sevilla, 41080 Seville, Spain. abandera@us.es, chacon@us.es} 
\and Cristina Caravaca García\thanks{Departamento de Ecuaciones Diferenciales y Análisis Numérico, Apdo. de correos 1160, Universidad de Sevilla, 41080 Seville, Spain. ccaravaca@us.es, edelgado1@us.es, macarena@us.es} 
\and Tomás Chacón Rebollo\footnotemark[1]  
\and Enrique Delgado Ávila\footnotemark[2]
\and Macarena Gómez Mármol\footnotemark[2]}

\date{}

\maketitle

\begin{abstract}

In this work, we introduce an \textit{a posteriori} error indicator for the reduced basis modelling of turbulent flows. It is based upon the $k^{-5/3}$ Kolmogorov turbulence theory, thus it may be applied to any numerical discretisation of LES turbulence models. The main idea of this indicator is that if the full-order solution and the Reduced Order solution are close enough, then their flow energy spectrum within the inertial range should also be close.  We present some numerical tests which supports that the use of this indicator is helpful, obtaining large computational speed-ups. We use as full-order model a Finite Element discretisation of the unsteady LES Smagorinsky turbulence model.

\end{abstract}

\paragraph{Keywords:}  Reduced Order Modelling, Large Eddy Simulation, Kol\-mo\-go\-rov energy cascade, Reduced Basis Method, Greedy algorithm.

\section{Introduction}

Reduced Order Modelling (ROM) has been successfully used in several fields to provide large reduction in computation times to solve Partial Differential Equations \cite{hestaven, holmes, LibroBR}. In fluid mechanics, a popular strategy is to use the Proper Orthogonal Decomposition (POD) \cite{POD1,POD_chapter} to extract the dominant structures for high-Reynolds flows, which are then used in a Galerkin approximation of the underlying equations \cite{holmes, sirovich}. Application of the POD-Galerkin strategy to turbulent fluid flows remains a challenging area of research \cite{busto2020pod,hijazi2023pod,hijazi2020data}. By construction, ROMs generated using only the first most energetic POD basis functions are not endowed with the dissipative mechanisms associated to the creation of lower size, and less energetic, turbulent scales. When the range of Reynolds is large, it is common to apply a Reduced Basis (RB) procedure to avoid the computation of a numerous amount of solutions of the original problem \cite{chacon2018assessment,fick2018stabilized,hijazi2020effort,Stokes2,Stokes1,Stokes4,Stokes3}. However, this sets the problem sampling, that could result in serious technical difficulties related to the building of error estimations for the Reduced Basis discretisation, as the ones developed in \cite{LPS2023, PaperSmago, Deparis, Dparis-Rozza}, based upon the Brezzi-Rappaz-Raviart theory \cite{BRR}. In this work, we present an \textit{a posteriori} error indicator, based upon the Kolmogorov turbulence theory, to overcome this problem of the RB procedure. 

Kolmogorov's energy cascade theory \cite{bouchon1998dynamic,TomasSmago,kolmogorov1941local,onsager1949statistical,richardson1922weather} is a fundamental concept in the field of fluid mechanics, and it provides a theoretical framework for understanding the energy transfer between different scales in a turbulent flow. In a fully developed turbulent flow, the kinetic energy is transferred from the large-scale eddies to smaller and smaller scales through a series of nonlinear interactions until it reaches the smallest scales, where it is dissipated into heat. This process is known as the energy cascade, and it is characterized by a self-similar scaling behaviour in the inertial subrange of scales. Kolmogorov's theory provides a statistical description of turbulence that has been widely used to guide the development of turbulence models and numerical simulations of turbulent flows.

In the context of ROM of Large Eddy Simulation (LES), since the full-order model (FOM) is intended to be a good approximation of the continuous problem, it should accurately approximate the energy spectrum of the continuous problem in the resolved part of the inertial spectrum. The key is to use the error deviation with respect to the full-order energy spectrum by the RB solution to select the new basis functions by the Greedy algorithm. This error indication procedure has the advantage of applying to any kind of numerical discretisation to both physical and geometrical parameters, and to any physical time at which the turbulence is in statistical equilibrium, as it is based on physical properties instead of model and discretization properties. 

To validate this indicator, we develop an academic test for 2D periodic flows exhibiting the $k^{-5/3}$ spectrum predicted by Kolmogorov's theory. In this test, we compare the use of the Kolmogorov indicator as error estimator with the use of the exact error between the full order and reduced order solution, as the basic criterion to select the basis functions in the Greedy algorithm. We address a turbulence model to be able to solve a part of the inertial spectrum. Actually, we consider an unsteady RB Smagorinsky turbulence model. The Smagorinsky model is a basic Large Eddy Simulation (LES) turbulence model \cite{Delgado2020,PaperSmago,TomasSmago,Sagaut,Smago}, that provides accurate solutions for the large scales of the flow, and a part of the inertial spectrum. Moreover, we use the Empirical Interpolation Method (EIM) \cite{EIM2,EIM1,MagicPoints} to build reduced approximations of the non-linear eddy viscosity term, coming from the Smagorinsky model. This allows to tensorise the eddy viscosity term, providing an efficient decoupling of the RB problem into an offline/online procedure. We obtain a speed-up of computing times nearly of 18, with error levels quite close to the optimal ones that would be obtained if the exact error is used as "error indicator" to build the reduced basis. Moreover, we observe a spectral decay of the error in terms of the dimension of the reduced space, quite close to the one obtained if the exact error is used as indicator.

The structure of this work is as follows. In Section \ref{sec:RB Kolmogorov}, we present the \textit{a posteriori} error indicator based upon the Kolmogorov turbulence theory and summarize the procedure to build the reduced basis model using this indicator. In Section \ref{sec:TestCase}, we present the unsteady Smagorinsky turbulence model. The numerical tests are presented in Section \ref{sec:Numerical_tests}, where we build a reduced basis model of the Smagorinsky turbulence model. Finally, Section \ref{sec:Conclusions} is devoted to the presentation of the conclusions of the results.

\section{Reduced Basis modelling of turbulent flows based on Kolmogorov’s theory}\label{sec:RB Kolmogorov}

The aim of this section is to introduce an \textit{a posteriori} error indicator for the selection of the parameter in the Greedy procedure for constructing the reduced basis model of a turbulent flow. This indicator is based upon the Kolmogorov turbulence theory, which introduces an expression for the energy cascade \cite{bouchon1998dynamic,TomasSmago,kolmogorov1941local,onsager1949statistical,richardson1922weather}. The main idea of this indicator is that a trial solution is accurate if its energy spectrum is close to the theoretical $k^{-5/3}$ spectrum predicted by the Kolmogorov theory. 

\subsection{\textit{A posteriori} error indicator based upon Kolmo\-gorov's theory}\label{sec:Kolmogorov}

Andrei Kolmogorov stated in \cite{kolmogorov1941local} that under suitable similarity and isotropy assumptions for turbulent flow in statistical equilibrium, there exists an inertial range $[k_1, k_2]$ where the energy spectrum $E(k)$ can be expressed in terms of the wavenumber $k$ and the turbulent dissipation $\varepsilon$, this is, $E(k)\sim \varepsilon^{2/3}k^{-5/3}$.

For a high fidelity solution 
we can assume that, in the inertial subrange, the energy spectrum accurately follows the expression
\begin{equation}\label{eq:energyKolmogorov}
E(k,\Re) = \alpha(\Re) k ^{-5/3},
\end{equation}
where $\alpha(\Re) >0$ depends on the turbulent dissipation $\varepsilon$ of the solution for the Reynolds number, $\Re$. Since the full-order model is intended to be a good approximation of the continuous problem, it should accurately approximate the energy spectrum of the continuous problem in the resolved part of the inertial spectrum, thus, it is fair to assume that the full-order model spectrum closely follows \eqref{eq:energyKolmogorov}.

Let us denote $E_N(k;\Re)$ be the energy spectrum associated to the reduced solution that belongs to a reduced space of dimension $N$. We define the \textit{a posteriori} error indicator as follows 
\begin{equation}\label{eq::Kolmogorov_apost_k2}
\Delta_N(\Re) = \min_{\alpha\in\R}\left( \int_{k_1}^{k_2}|E_N(k;\Re) - \alpha(\Re)k^{-5/3}|\, dk \right)^{1/2}.
\end{equation}
This indicator measures how close is a given ROM solution to the theoretical Kolmogorov spectrum in the inertial range $[k_1,k_2]$. The main advantage is that it can be applied no matter the numerical discretization and the turbulence model we are working with, as it only involves the spectrum of the trial solution. 

This allows us to overcome the high technical difficulties related to the construction of a suitable \textit{a posteriori} error estimator for the Reduced Basis discretization, as the ones developed in \cite{LPS2023, PaperSmago, Deparis, Manzoni}, based upon the Brezzi-Rappaz-Raviart theory \cite{BRR} and that necessarily are specific for each actual turbulence model and numerical discretisation.

In practice, we substitute $k_2$ by $k_c<k_2$, where $k_c$ is the smallest scale we can solve numerically by means of the turbulence model considered. In most cases, $k_c$ is related to the mesh size $\delta$. Since this indicator is valid whenever the turbulence model solves a range of the inertial spectrum, the mesh $\Th$ should be carefully chosen in order to effectively solve a part of the inertial range, that is, $k_c\in(k_1,k_2)$, in practice $k_c\ll k_2$. 

\subsection{General algorithms to build the reduced basis spaces}\label{sec:General algorithm}

In this section, we describe a general algorithm based on the \textit{a posteriori} error indicator $\Delta_N(\Re)$, presented in \eqref{eq::Kolmogorov_apost_k2}, for the construction of the reduced basis model of LES turbulent model. We describe it here in general, as it is independent of the actual LES model considered. We particularize it to the Smagorinsky turbulence model in the next section. 

We intend to solve parametric turbulent flows depending on a parameter $\mu\in\R^p$, for any integer $p\ge1$. In the applications that follow this parameter will be the Reynolds number, although it can also be a different physical or geometrical parameter, o several of these. We shall suppose that $\mu$ ranges on a compact set $\cD\subset\R^p$.

The general algorithm for the basis functions selection can be summarized as follows:

\begin{enumerate}
    \item First, we determine a partition, $\cD_{train}$, of the parameter space, $\cD$, such that $\mathcal{D}_{train} \subset \mathcal{D}$. We also need to set the stopping criteria, and select an initial parameter $\mu^*_0\in \mathcal{D}
    _{train}$.
    \item For a given reduced space, solve the full-order turbulence model at the different time steps (called ``snapshots"), for a certain time range $[t_1,t_L]$ and store the results of the velocity and pressure. It is needed that at the initial time, $t_1$ the turbulence is already at statistical equilibrium.
    \item Apply some technique to compress the snapshots for the velocity and pressure, and add them to the previous velocity and pressure Reduced Basis spaces. 
    \item Compute the \textit{a posteriori} error indicator $\Delta_N(\mu)$ for each $\mu\in \mathcal{D}_{train}$ associated to the reduced basis. The error indicator should be computed for some discrete time sets to be determined. 
    \item Select the new parameter $\mu^*$ according to some criterium based upon the error indicator $\Delta_N$ (see Subsection \ref{sec:SelectionCriteria} below).
    \item Check the stopping criteria. If the criteria are fulfilled, then stop the algorithm, otherwise, repeat from step 2.
\end{enumerate}



\subsection{Criteria for the selection of new parameter values}\label{sec:SelectionCriteria}

Since the reduced solution is built from the FE approximation, we should not expect that $\Delta_N(\mu)$, defined in \eqref{eq::Kolmogorov_apost_k2}, tends to $0$ as $N$ grows, it should rather converge to $\Delta_h(\mu)$, with 
$$\Delta_h(\mu)=\min_{\alpha\in\R}\left(\int_{k_1}^{k_c}|E_h(k;\mu)-\alpha(\mu)k^{-5/3} |^2\ dk\right)^{1/2}$$
where $E_h(k;\mu)$ for $k\in(k_1,k_c)$ represents the energy spectrum of the full-order solution. This spectrum should be close to the theoretical $k^{-5/3}$ law if the FOM is accurately but, however, some error is to be expected as the FOM is not exact.

Then, the usual criterium for the selection of the best parameter in the Greedy algorithm, that is $\mu^*=\max_{\mu\in\cD}\Delta_N(\mu)$, is no longer useful. Instead, we propose in Algorithm \ref{alg:SelectionParameter} new criteria for the selection of the next pa\-ra\-me\-ter at each step. Algorithm \ref{alg:SelectionParameter} first aims to add to the basis the information of the solutions whose energy spectrum is farther away from the theoretical one, then, if the parameter selected that way has been already selected in a previous iteration, we select between the non-selected parameters, the one that provides the largest difference between the energy spectrums of the previous and the current reduced solutions. In the latter case, we are assuming that the reduced error indicator $\Delta_N(\mu)$ is getting close to $\Delta_h(\mu)$, and the previous solution plays the role of the FOM one to compare with it. 

We will compute the value of the indicator at the final time $t_L$, since the error with respect to the FOM solution is expected to increase as time increases.

\begin{algorithm}
\caption{Parameter selection criteria}\label{alg:SelectionParameter}
\begin{algorithmic}
\State Let $S_{N}$ be the set of previously selected parameters. 
\State Compute $\mu^*_N=\arg\max_{\mu\in\cD{train}}\Delta_N(\mu,t_L)$;
\If{$\mu^*_N\in S_{N}$}
    \State $\mu^*_N = \arg\max_{\mu\in\cD{train}\setminus S_{N}}|\Delta_N(\mu,t_L)-\Delta_{N-1}(\mu,t_L)|$;
\EndIf
\end{algorithmic}
\end{algorithm}

\subsection{Determination of new reduced space}\label{sec:reduced_space_construction} %

In this section, we summarize the procedure used for the construction of the reduced spaces, actually through a POD+Greedy approach \cite{Haasdonk2013}. 

We follow a POD strategy considering the time as a parameter, and the Greedy algorithm for the physical parameter. For the POD, we use a separate strategy in the sense that we apply the POD to a velocity snapshot set, and also to a pressure snapshot set. That is, we solve the high-fidelity problem for a parameter value $\mu^*$, and we perform a POD, for velocity and pressure separately, for the selected time snapshots. Then, we select the next parameter value $\mu^*$ with the \textit{a posteriori} error indicator criteria. 

This procedure is summarised in the following:
\begin{enumerate}
    \item For a given $\mu^*$, solve the Smagorinsky Model for any discretisation time $t_n$ for $n=1,\dots,L$, and add the time snapshots for velocity and pressure to the reduced velocity and pressure spaces, $Y_{N-1}$ and $M_{N-1}$ respectively, obtaining the new reduced velocity and pressure spaces ${Y}_N$ and ${M}_N$, respectively.
    \item Apply a POD to the reduced velocity and pressure spaces ${Y}_N$ and ${M}_N$, for a given tolerance $\varepsilon$. 
    \item Compute the inner pressure \textit{supremizer} for the pressure basis resulting above (\textit{cf.} \cite{Ballarin}), and we add it to the reduced velocity space (see Subsection \ref{sec:RBP} for a detailed description). This step is needed in order to guarantee the inf-sup condition for the pair of reduced valocity and pressure spaces. If some stabilization techniques are considered (see e.g. \cite{Ali2020, LPS2023}) this step would be no longer needed to be taken into account.
    \item Finally, apply the selection criteria presented in Algorithm \ref{alg:SelectionParameter} for the RB problem associated to the spaces $Y_N$ and $M_N$, obtaining the new parameter $\mu^*$. 
\end{enumerate}

We could also use a similar procedure as the one presented in \cite{caravaca2022reduced}, where two POD procedures were performed. In step 1, it can be possible to perform separate POD reductions for the time snapshots for velocity and pressure. Then, the basis obtained by this POD are added to the reduced velocity and pressure spaces. Then, we follow the step 2, performing again a POD to the reduced velocity and pressure spaces. We will compare the results obtained for both procedures in Section \ref{sec:Numerical_tests}.

\section{Smagorinsky turbulence model}\label{sec:TestCase}

This section is devoted to the construction of the reduced LES turbulence model that we shall use to test the error indicator introduced in Section \ref{sec:RB Kolmogorov}. Actually,  we present the unsteady Smagorinsky turbulence model, that is the basic LES turbulence model, in which the effect of the subgrid scales on the resolved scales is modelled by eddy diffusion terms \cite{TomasSmago}. We introduce a discretisation by the Finite Element method using inf-sup stable velocity-pressure spaces.

\subsection{Finite element problem}\label{sec:HighFid}

Let $\Omega$ be a bounded domain of $\R^d\, (d=2,3)$, with Lipschitz-continuous boundary $\Gamma$. 
We present a parametric unsteady Smagorinsky turbulence model, where in the following we consider the Reynolds number as a phy\-si\-cal parameter, denoting it by $\mu\in\cD$, where $\cD$ is an interval of positive real numbers, large enough to ensure that the flow is in turbulent regime. Also, we consider the time interval $[0, T_f]$, with $T_f >0$ a chosen finite time. Let $\{\Th\}_{h>0}$ be a uniformly regular mesh in the sense of Ciarlet \cite{Ciarlet1978}.  We are denoting by $h_K$ the diameter of an element $K\in\Th$. In order to clarify the relationship between the Smagorinsky model and the Navier-Stokes equations, we present the model as a continuous one, although it is intrinsically discrete. We search for a velocity field $\uk \,: \overline{\Omega}\times [0,T_f] \mapsto \R^d$ and a pressure function $p\,:\overline{\Omega} \times [0,T_f] \mapsto \R$ such that
\begin{equation}\label{NS}\left\{\begin{array}{ll}
\partial_t\uk + \uk\cdot\nabla\uk+\nabla p-\nabla\cdot\left(\left(\dfrac{1}{\mu}+\nu_T(\uk)\right)\nabla\uk\right)=\fk&\mbox{ in }\Omega\times [0,T_f],\vspace{0.2cm}\\

\nabla\cdot\uk={\bf 0}&\mbox{ in } \Omega\times [0,T_f],\vspace{0.1cm}\\
\end{array}\right.
\end{equation}
plus boundary and initial conditions, where $\fk$ is the kinetic momentum source, and $\nu_T(\uk)$ is the eddy viscosity defined as
\begin{equation} \label{eq:eddivis}
\nu_T(\uk) = C_S^2\D\sum_{K\in\mathcal{T}_h}h_K^2\big|\nabla\uk_{|_K}\big|\chi_K, 
\end{equation} 
where $\big|\cdot\big|$ denotes the Frobenius norm in $\R^{d\times d}$, and $C_S$ is the Smagorinsky cons\-tant.  

To state the full-order discretisation that we consider for problem \eqref{NS}, let us introduce the velocity and pressure spaces
$$
Y=\{ \vk \in H^1(\Omega)^d,\,\, \mbox{s.t. } \vk_{|_{\Gamma_D}}={\bf 0}\,\},\quad M=\{q \in L^2(\Omega),\,\, \mbox{s.t. }\, \int_\Omega q =0\,\}.
$$
We assume $\fk \in Y'$.

Let $Y_h$ and $M_h$ be two finite element subspaces of $Y$ and $M$, respectively, that satisfy the discrete inf-sup condition, i.e.,
\begin{equation}\label{infsupFE}
\normld{q_h}=\sup_{\vk_h\in Y_h}\dfrac{(q_h,\nabla\cdot\vk_h)_\Omega}{\normld{\nabla{\vk_h}}}, \quad \forall q_h\in M_h.
\end{equation}
 We consider the following Galerkin discretisation of the unsteady Smagorinsky model \eqref{NS}, 
\begin{equation}
\label{chap:VMS::pb:FV}\left\{\begin{array}{l}
\forall\mu\in\cD \text{ and } t\in[0, T_f], \mbox{find } (\uk_h(t;\mu),p_h(t;\mu))\in Y_h\times M_h\mbox{ such that}\vspace{0.3cm}\\
\begin{array}{ll}
(\partial_t\uk_h, \vk_h)_\Omega + a(\uk_h,\vk_h;\mu)+b(\vk_h,p_h;\mu)\\
+\, a_S(\uk_h;\uk_h,\vk_h;\mu) + c(\uk_h,\uk_h,\vk_h;\mu)=\left<\fk,\vk_h\right>&\quad\forall\vk_h\in Y_h,\vspace{0.1cm}\\
b(\uk_h,q_h;\mu)=0&\quad\forall q_h\in M_h,\end{array}\end{array}\right.
\end{equation} 
where the bilinear forms $a(\cdot,\cdot;\mu)$ and $b(\cdot,\cdot;\mu)$ are defined as 
\begin{equation}\label{chap:Smago::eq:ab_form}
a(\uk,\vk;\mu)=\frac{1}{\mu}\int_\Omega\nabla\uk:\nabla\vk\,d\Omega,\qquad b(\vk,q;\mu)=-\int_\Omega(\nabla\cdot\vk)q\,d\Omega;
\end{equation}
while the trilinear form $c(\cdot,\cdot,\cdot;\mu)$ is defined as
\begin{equation}\label{chap:Smago::eq:conv_form}
c(\zk,\uk,\vk;\mu)=\frac{1}{2}\, \left [\int_\Omega(\zk\cdot\nabla\uk)\vk\,d\Omega- \int_\Omega(\zk\cdot\nabla\vk)\uk\,d\Omega\, \right ].
\end{equation}
Moreover, the non-linear form  $a_S(\cdot;\cdot,\cdot;\mu)$, is a Smagorinksy mo\-de\-lling for the eddy viscosity term, and it is given by
\begin{equation}\label{chap:VMS::eq:Small-Small}
a_S(\zk;\uk,\vk;\mu)=\intO{\nu_T(\zk)\;\nabla\uk:\nabla\vk}.
\end{equation}

If the inf-sup condition \eqref{infsupFE} is satisfied by the Finite Element pair of velocity-pressure spaces $Y_h$ and $M_h$, problem \eqref{chap:VMS::pb:FV} is well-possed. See \cite{TomasSmago} for more details. 

\subsection{Reduced basis problem}\label{sec:RBP}
In this section, we introduce the Reduced Basis (RB) model for problem \eqref{NS}. The RB problem also is a Galerkin projection of the Smagorinsky model, but now on the reduced spaces. It reads 
\begin{equation}\label{chap:VMS::pb:RBP}\left\{\begin{array}{l}
\forall\mu\in\cD \text{ and } t\in[0, T_f], \mbox{find } (\uk_N(t;\mu),p_N(t;\mu))\in Y_N\times M_N\mbox{ such that}\vspace{0.3cm}\\

\begin{array}{ll}
    (\partial_t\uk_N, \vk_N)_\Omega + a(\uk_N,\vk_N;\mu)+b(\vk_N,p_N;\mu) \\
    + a_S(\uk_N;\uk_N,\vk_N;\mu) +c(\uk_N,\uk_N,\vk_N;\mu)=\langle \fk, \vk_N\rangle &\quad\forall\vk_N\in Y_N,\vspace{0.1cm}\\
b(\uk_N,q_N;\mu) =0&\quad\forall q_N\in M_N.\end{array}\end{array}\right.
\end{equation} 

Here, %
we recall that we denote by $Y_N$ the reduced velocity space, and by  $M_N$ the reduced pressure space. Their dimensions are intended to be much smaller than their finite element counterparts, $Y_h$ and $M_h$. The computation of the reduced spaces is done following the POD+Greedy approach explained in Section \ref{sec:reduced_space_construction}, and described in Algorithm \ref{alg:basic:1PODgreedySupremizer}.

\begin{algorithm}
\caption{POD+Greedy with \textit{supremizer}}\label{alg:basic:1PODgreedySupremizer}
\begin{algorithmic}
\State Set $\epsilon_{tol}, \epsilon_{POD}>0$, $N_{max}\in\mathbb{N}$, $\mu^*_0\in\cD_{train}$, $\mathbb{Z}^u=[\ ]$, $\mathbb{Z}^p=[\ ]$, $N=0$, and $S^*=\{\mu_0^* \}$;
\While{$\cN < N_{\max}$}
    \State Compute $(\uk_h^n(\mu^*_N),p_h^n(\mu^*_N))$ for $n=1,\dots,L$;
    \State $\mathbb{Z}^u=[\mathbb{Z}^u,\uk_h^1(\mu^*_N),\uk_h^2(\mu^*_N),\dots,\uk_h^{L}(\mu^*_N)]$;
    \State $\mathbb{Z}^p=[\mathbb{Z}^p, {p}_h^1(\mu^*_N),{p}_h^2(\mu^*_N),\dots,{p}_h^{L}(\mu^*_N)]$;
    \State $[{\varphi}^u_1,\dots,{\varphi}^u_{\cN^u}]$ = POD($\mathbb{Z}^u,\epsilon_{POD}$);
    \State $[{\varphi}^p_1,\dots,{\varphi}^p_{\cN^p}]$ = POD($\mathbb{Z}^p,\epsilon_{POD}$);
    \State Compute $\varphi_{\cN^u+i}^u = T_p^u\varphi_i^p$ for $i = 1,\dots,\cN^p$;
    \State $\cN=\cN^u+2\cN^p$;
    \State $Y_N=\{\varphi^u_i\}_{i=1}^{\cN^u+\cN^p}$, $M_N=\{\varphi^p_i\}_{i=1}^{\cN^p}$, 
    \State $\mathbb{Z}^u=[\varphi^u_1,\dots,\varphi^u_{\cN_u+\cN_p}]$, $\mathbb{Z}^p=[\varphi^p_1,\dots,\varphi^p_{\cN_p}]$
    \State Apply Algorithm \ref{alg:SelectionParameter} for selecting the parameter value $\mu^*_{N+1}$.
    \State $\epsilon_{N}=\Delta_{N}(\mu^*_{N+1},t_L)$;    
    \If{$\epsilon_{N}\leq \epsilon_{tol}$}
        \State $N_{max}=\cN$;
    \EndIf
    \State $S_{N+1}=S_{N}\cup \{\mu^*_{N+1}\}$ and $N = N + 1$;
\EndWhile
\end{algorithmic}
\end{algorithm}

To build the POD correlation matrices, for velocity and pressure, it is necessary to establish spatial norms for velocity and pressure spaces, since the time should be considered as a parameter. In this case, we use the $H^1$-seminorm, and $L^2$-norm, for the spaces $Y_h$ and $M_h$, respectively. 

In order to guarantee the inf-sup condition \eqref{infsupFE} for the reduced spaces, we use the so-called inner pressure \textit{supremizer}  (see e.g. \cite{Ballarin, Stabile2018}), defined by 
\begin{equation}\label{supremizer_def}
T_p^\mu\in Y_n \text{ such that }\left(\nabla T_p^\mu q_h,\nabla \vk_h\right)_\Omega=b(q_h,\vk_h;\mu)\quad\forall\vk_h\in Y_h.
\end{equation}

With the consideration of the inner pressure \textit{supremizer} for the velocity space,  the pair of reduced velocity and pressure spaces satisfies an equivalent inf-sup condition of \eqref{infsupFE}. Thus, probem \eqref{chap:VMS::pb:RBP} is also well-posed (see eg. \cite{TomasSmago}).

The Smagorinsky eddy viscosity term defined in \eqref{eq:eddivis}, $\nu_T(\nabla\uk):=\nu_T(\mu)$, is a non-linear function of the parameter, and consequently needs to be tensorised in the offline phase. We use the Empirical Interpolation Method (EIM) \cite{EIM2, EIM1,MagicPoints} to build the RB model, in order to reduce the online computation times. 

For this purpose, we build a reduced-basis space for the eddy viscosity, $W_{M}^S=\{q_1^S(\mu),\dots,q_{M_1}^S(\mu)\}$ by a greedy procedure selection. In this case, we consider the time also as a parameter jointly with the Reynolds number. Thus, we approximate the non-linear Smagorinsky term by the following trilinear form:
\begin{equation}
a_S(\uk_N;\uk_N,\vk_N;\mu)\approx \hat{a}_S(\uk_N,\vk_N;\mu),
\end{equation}
where,
\begin{equation}
\hat{a}_S(\uk_N,\vk_N;\mu)=\dsum_{k=1}^{M_1}\sigma^S_k(\mu)s(q_k^S,\uk_N,\vk_N),
\end{equation}
with,
\begin{equation}
s(q_k^S,\uk_N,\vk_N)=\dsum_{K\in\Th}\big(q^S_k \,\nabla \uk_N,\nabla\vk_N\big)_K.
\end{equation}

In practise, we solve problem \eqref{chap:VMS::pb:RBP} considering $\hat{a}_S(\cdot,\cdot;\mu)$ instead of \\ $a_S(\cdot;\cdot,\cdot;\mu)$.


\section{Numerical tests}\label{sec:Numerical_tests}

In this section, we present the numerical tests of the practical performances of the estimator $\Delta_N$. In a first test, we consider a POD+Greedy procedure, where we select the parameter values with the Greedy algorithm. In a second test, we consider the parameter values as an equispaced partition of the parameter set to compare the results of both approaches.

We solve the Smagorinsky model  \eqref{chap:VMS::pb:FV} for 2D periodic flows, in the time interval $t\in[0,30]$, over the unit square $\Omega=[-1/2, 1/2]^2$ with periodic boundary conditions. We do not consider any source, thus $\fk=0$. We consider a structured mesh, where we divide each edge in $m=64$ intervals, obtaining a mesh with 8192 triangles and 4225 vertices. We use the inf-sup stable Taylor-Hood Finite Element, i.e., $\mathbb{P}2-\mathbb{P}1$ Finite Element for velocity-pressure discretisation. We also use a Crank-Nicolson scheme for the time derivative discretisation. We select the Reynolds number, $\mu$, as the parameter, ranging on $\cD=[1000, 16000]$. For this range, the flow reaches the Kolmogorov spectrum profile roughly at times larger than 10. The possible $k^{-3}$ inertial spectrum of 2D turbulence does not take place, as there is no forcing at large wavenumbers.

\subsection{Data of the problem and initial condition}

To determine the initial condition, we look for a velocity field with an inertial energy spectrum as in \eqref{eq:energyKolmogorov}. We consider a velocity field $\uk_h^0 = (v,v)$, where $v$ is defined through its Fourier transform:
\begin{equation}\label{eq:Kolmogorov_indicator}
\hat{v}(k) = \left\{\begin{array}{ll} k^{-(5/3+1)/2} & \text{if } 0<k\le m/4,\\0 & \text{other case.}
\end{array}\right.
\end{equation}

To determine the initial condition for problem \eqref{chap:VMS::pb:FV}, we solve the Smago\-rin\-sky model taking $\uk^0_h$ as the initial state for $\mu  = 8500$, the intermediate Reynolds number, and we take as initial condition the velocity field at $t=15$.  

\begin{figure}
 \centering
\begin{subfigure}{0.45\textwidth}
    \centering
    \includegraphics[width=\textwidth]{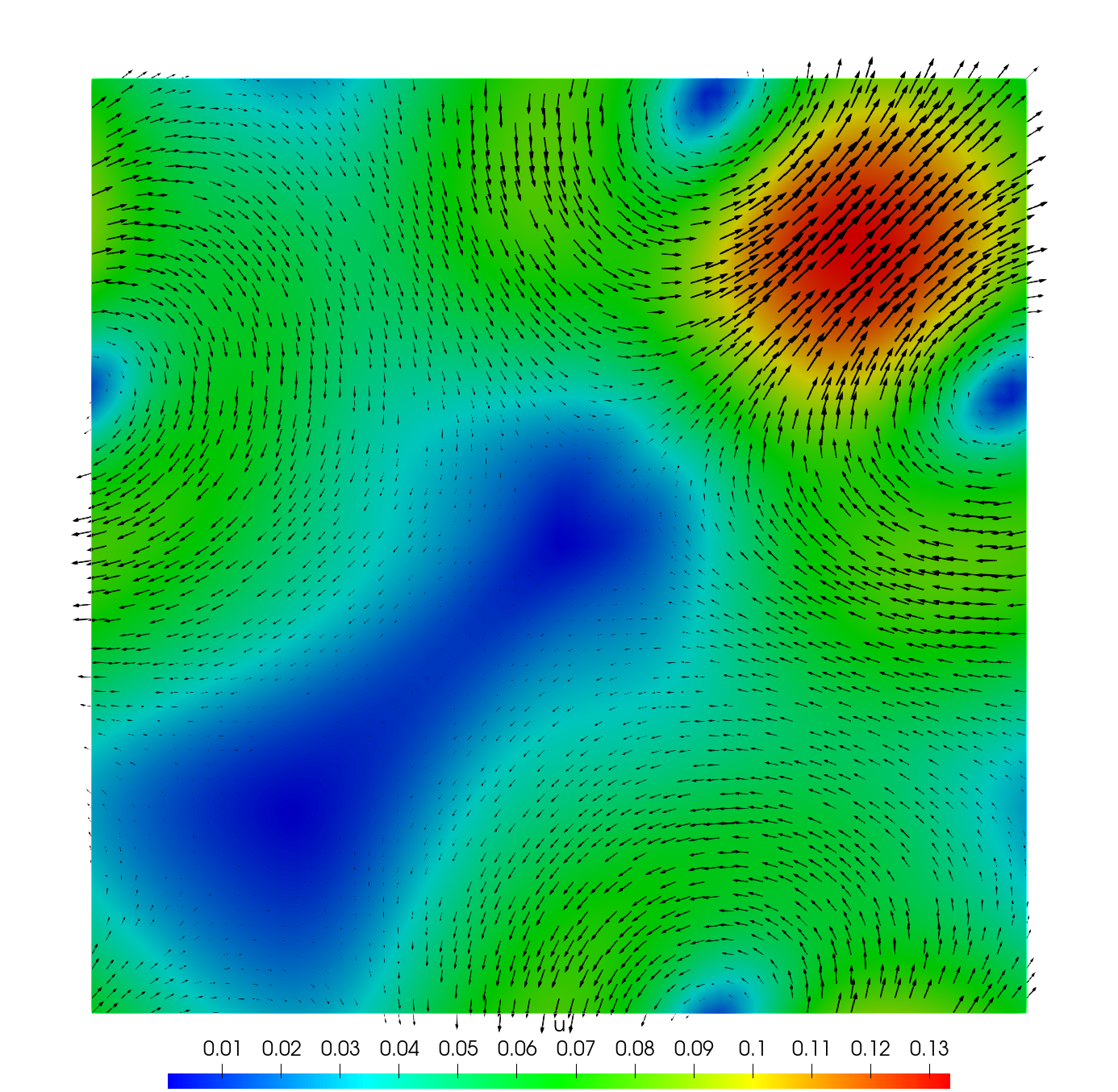}
\end{subfigure}
\begin{subfigure}{0.49\textwidth}
    \centering
    \includegraphics[width=\textwidth]{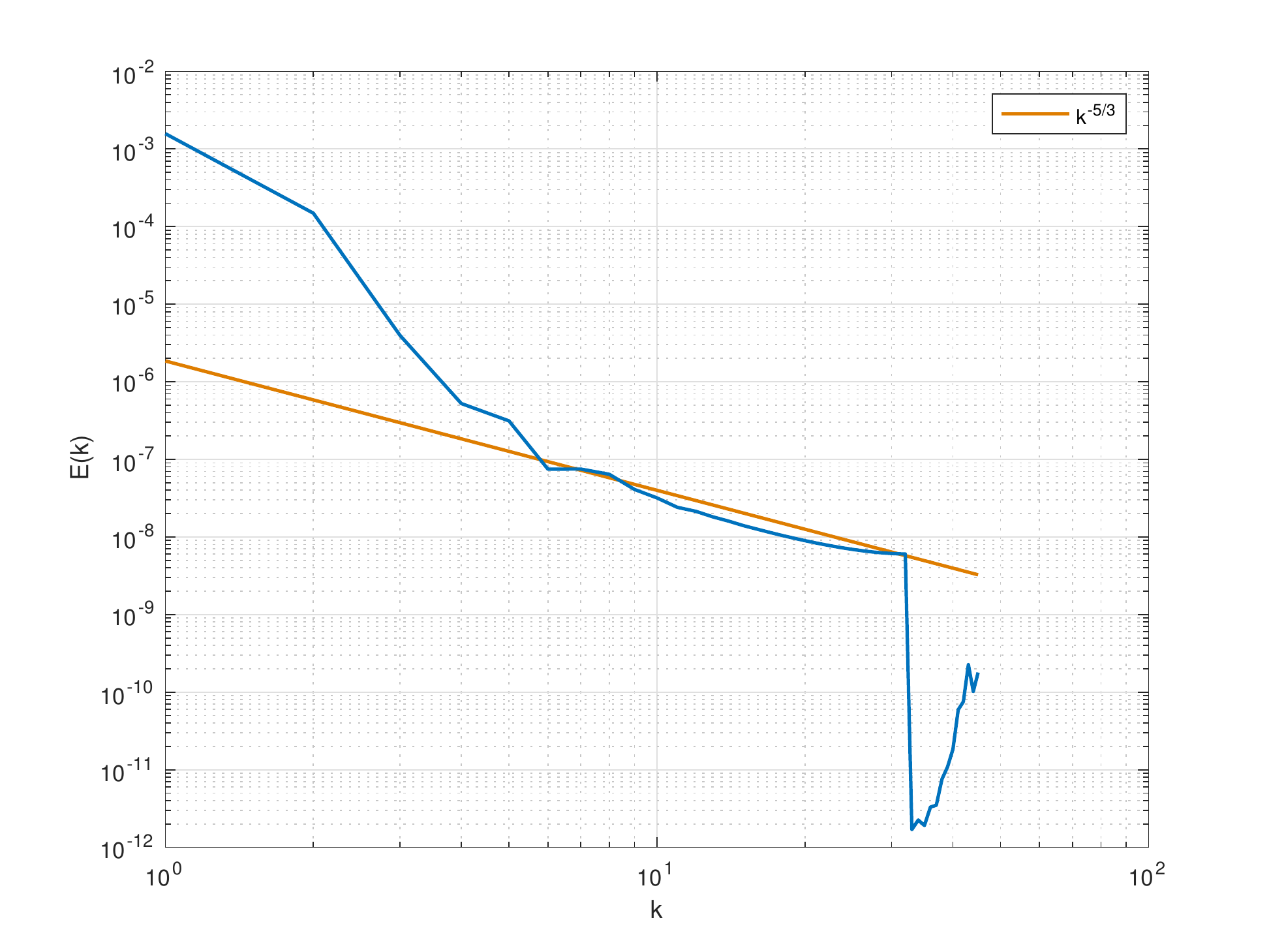}
\end{subfigure}
    \caption{Initial condition $\uk_h^0$ and its energy spectrum.}
    \label{fig:initialcondition}
\end{figure}

In Figure \ref{fig:initialcondition}, we show the initial condition $\uk_h^0$ and its energy spectrum. For wavenumbers between $k_1=5$ and $k_c=32$, we obtain a good approximation of the inertial spectrum. For $k>32$, we observe an abrupt decay of the energy. This is produced by the wavenumbers that are out of the circle of the largest radio inside the unit square and the viscous effects. As we can see, we start from a well-developed inertial energy spectrum.

\begin{figure}
 \centering
\begin{subfigure}{0.49\textwidth}
    \centering
    \includegraphics[width=\textwidth]{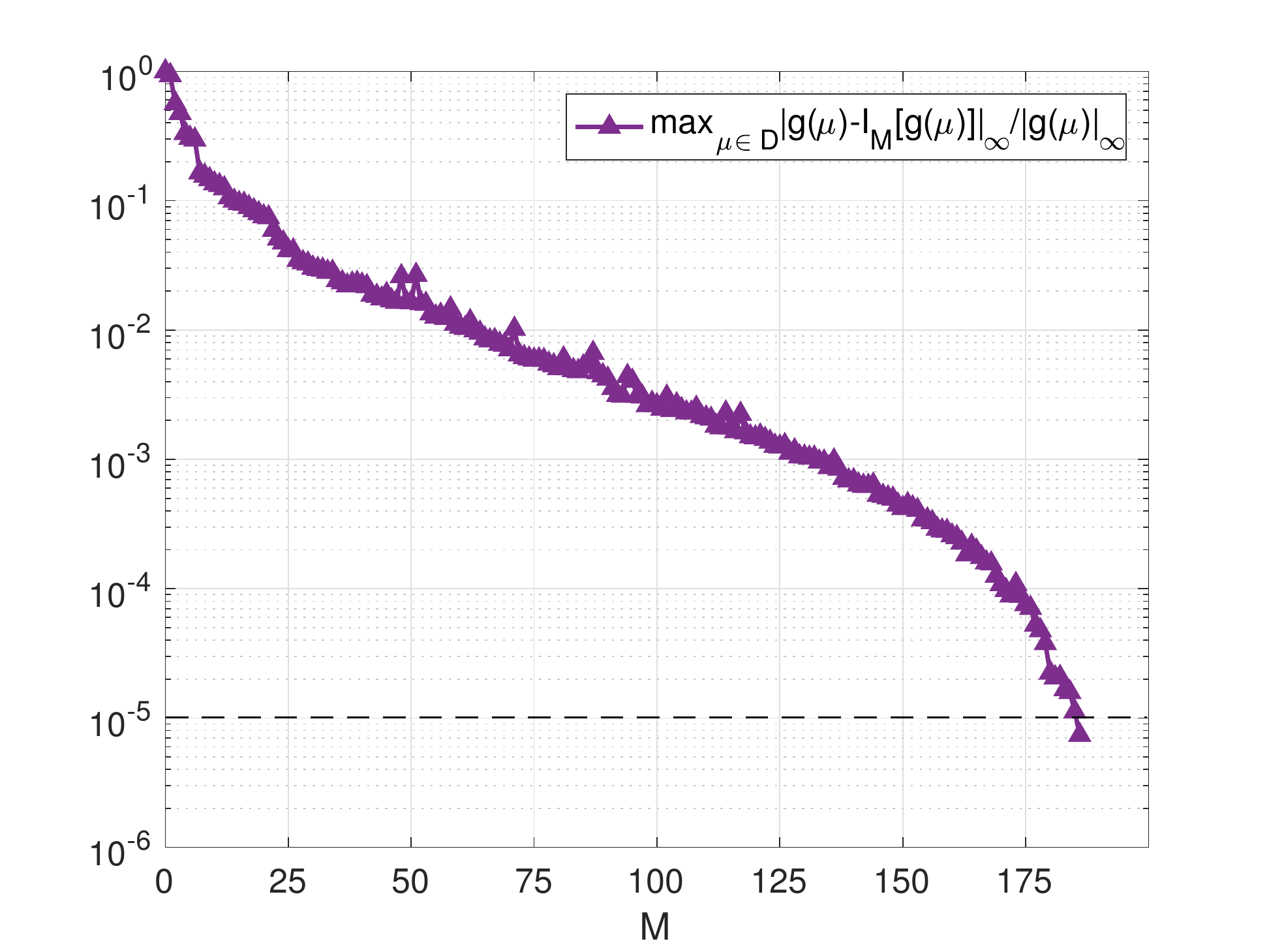}
    \caption{EIM history convergence error.}
    \label{fig:EIMHistoryKolmo}
\end{subfigure}
\begin{subfigure}{0.49\textwidth}
    \centering
    \includegraphics[width=\textwidth]{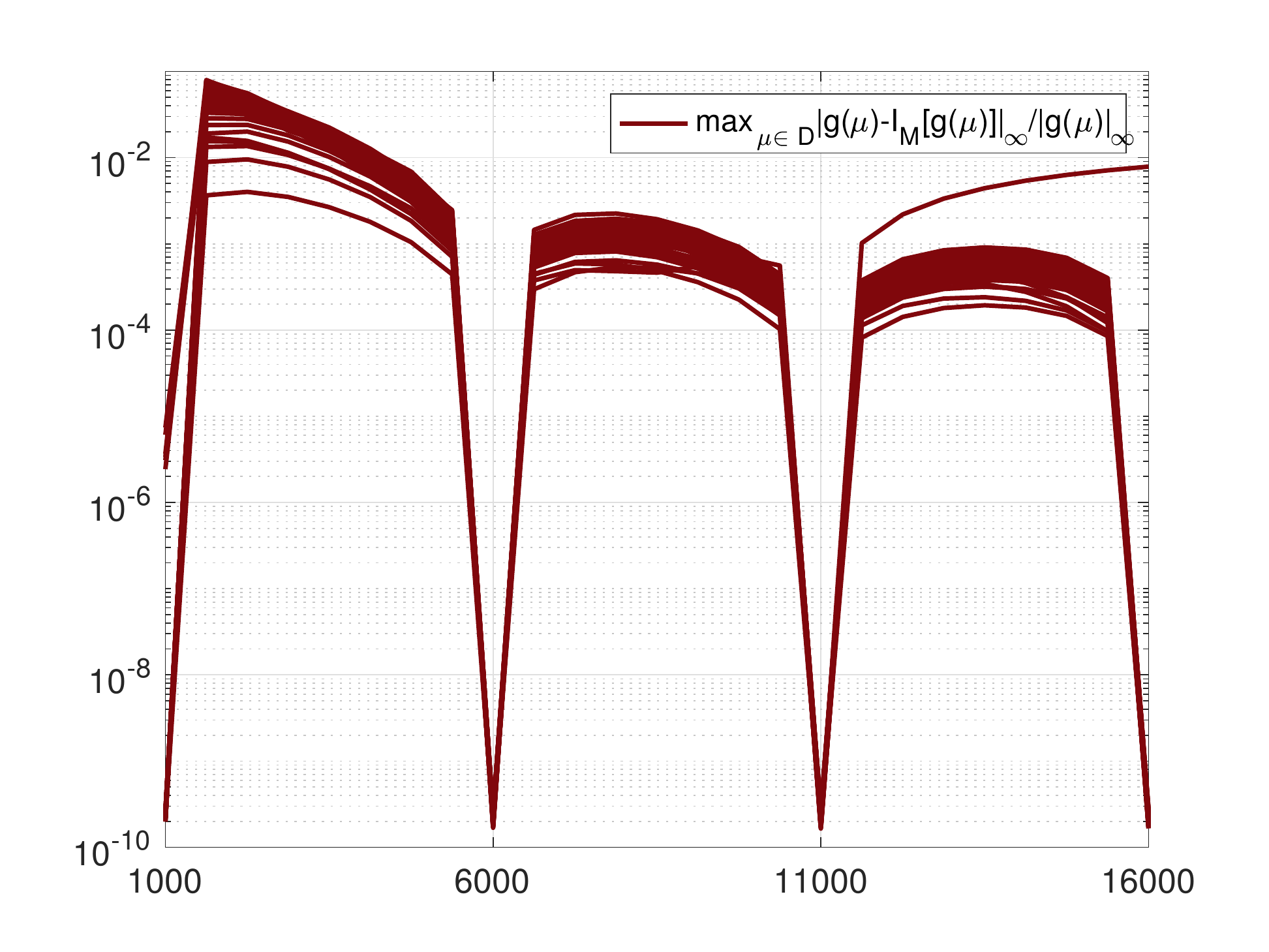}
    \caption{EIM error, each line corresponds to a time step $t_n$ for $n=1,\dots,L$.}
    \label{fig:EIMerrorKolmo}
\end{subfigure}
\caption{EIM applied to unsteady Smagorinsky Model.}
\end{figure}

As we mentioned, we need to tensorise the Smagorinsky eddy viscosity term \eqref{eq:eddivis}, with respect to the parameter. For this purpose, we compute the finite element solution $(\uk_h^n(\mu),p_h^n(\mu))$ for all $n=1,\dots,L$, and for $\mu=\{1\,000,\,6\,000,\,11\,000,\,16\,000\}$, we apply the EIM to compute the approximation of the eddy viscosity function. We stop the algorithm on $186$ basis functions when the error is below $\epsilon_{EIM}=10^{-5}$. The convergence error is shown in Figure \ref{fig:EIMHistoryKolmo}, while in Figure \ref{fig:EIMerrorKolmo}, we show the error, where each line represents a time step $t_{l}$ for $n=1,\dots,48$. 

\subsection{POD+Greedy algorithm}

To perform a first test of the \textit{a posteriori} error indicator, we compare the use of the indicator $\Delta_N(\mu)$ introduced in \eqref{eq::Kolmogorov_apost_k2} versus the use of the exact error at the final time, $T_f=30$, for the parameter selection in the Greedy algorithm. The latter provides the best possible choice of the new parameter.

\begin{table}[htb]
    \centering
    \begin{tabular}{|c|c|c|ccc|}
        \hline
        It. & $\mu$ & N & $\max_{\mu}\Delta_N(\mu)$ & $\max_{\mu}|\Delta_N(\mu)-\Delta_{N^*}(\mu)|$ & $l_2(\epsilon_N(\mu))$ \\ \hline
        1   & 1000  & 30  & $7,92\cdot 10^{-1}$ & & $2,32 \cdot 10^{ 0}$ \\ 
        2   & 16000 & 70  & $5,16\cdot 10^{-1}$ & & $1,99 \cdot 10^{-1}$ \\ 
        3   & 1625  & 91  & $3,53\cdot 10^{-1}$ & & $7,18 \cdot 10^{-2}$ \\ 
        4   & 2250  & 107 & & $3,93\cdot 10^{-2}$ & $2,22 \cdot 10^{-2}$ \\ 
        5   & 2875  & 117 & & $1,79\cdot 10^{-2}$ & $1,16 \cdot 10^{-2}$ \\ 
        6   & 6000  & 134 & & $2,08\cdot 10^{-3}$ & $2,53 \cdot 10^{-3}$ \\ 
        7   & 12875 & 145 & & $1,53\cdot 10^{-3}$ & $6,04 \cdot 10^{-4}$ \\ 
        8   & 10375 & 149 & & $2,23\cdot 10^{-4}$ & $4,29 \cdot 10^{-4}$ \\ \hline
    \end{tabular}
    \begin{tabular}{|c|c|c|c|}
        \hline
        It. & $\mu$ & N & $\max_{\mu} \epsilon_N(\mu)$ \\ \hline
        1   & 1000  & 30  & $5,54\cdot 10^{-1}$ \\ 
        2   & 16000 & 70  & $6,90\cdot 10^{-2}$ \\ 
        3   & 3250  & 96  & $1,34\cdot 10^{-2}$ \\ 
        4   & 1625  & 111 & $4,76 \cdot 10^{-3}$ \\ 
        5   & 8500  & 132 & $7,54 \cdot 10^{-4}$ \\ 
        6   & 5375  & 141 & $3,59 \cdot 10^{-4}$ \\ 
        7   & 12875 & 147 & $2,78 \cdot 10^{-4}$ \\ 
        8   & 2250  & 150 & $1,23 \cdot 10^{-4}$ \\ \hline
    \end{tabular}
    \caption{Convergence story of the POD+Greedy algorithm with one POD, using $\Delta_N(\mu)$, the indicator, (upper table) and $\epsilon_N(\mu)$, the exact error, (lower table) for the parameter selection.}
    \label{tab:PODgreedy1}
\end{table}

\begin{figure}
    \centering
    \includegraphics[width=0.8\textwidth]{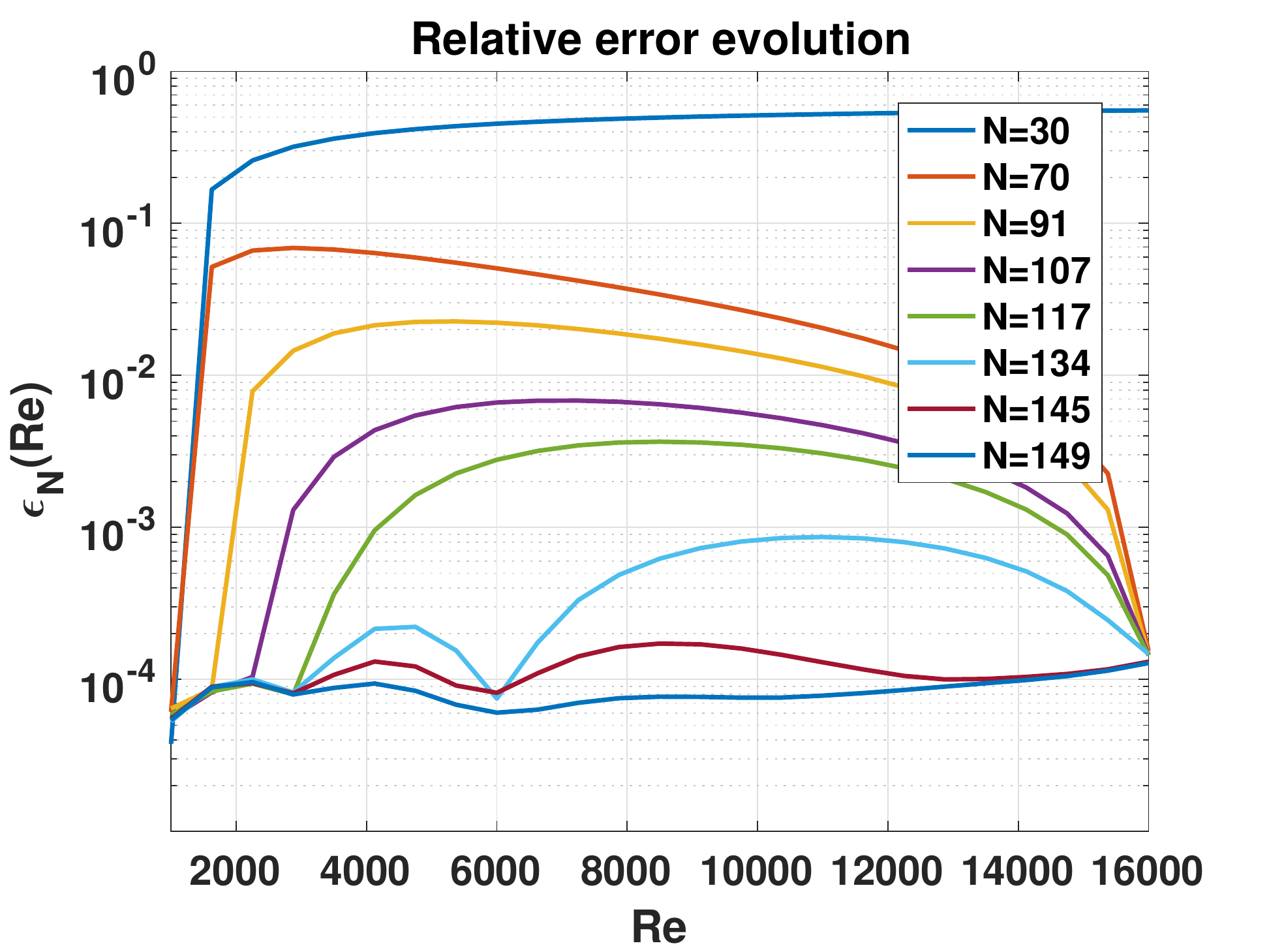}
    \caption{Error $\epsilon_N(\mu)$ in each iteration of the POD+Greedy algorithm, using the Kolmogorov error indicator.}
    \label{fig:errorEvolution 1POD}
\end{figure}

\begin{figure}
    \centering
    \includegraphics[width=0.8\textwidth]{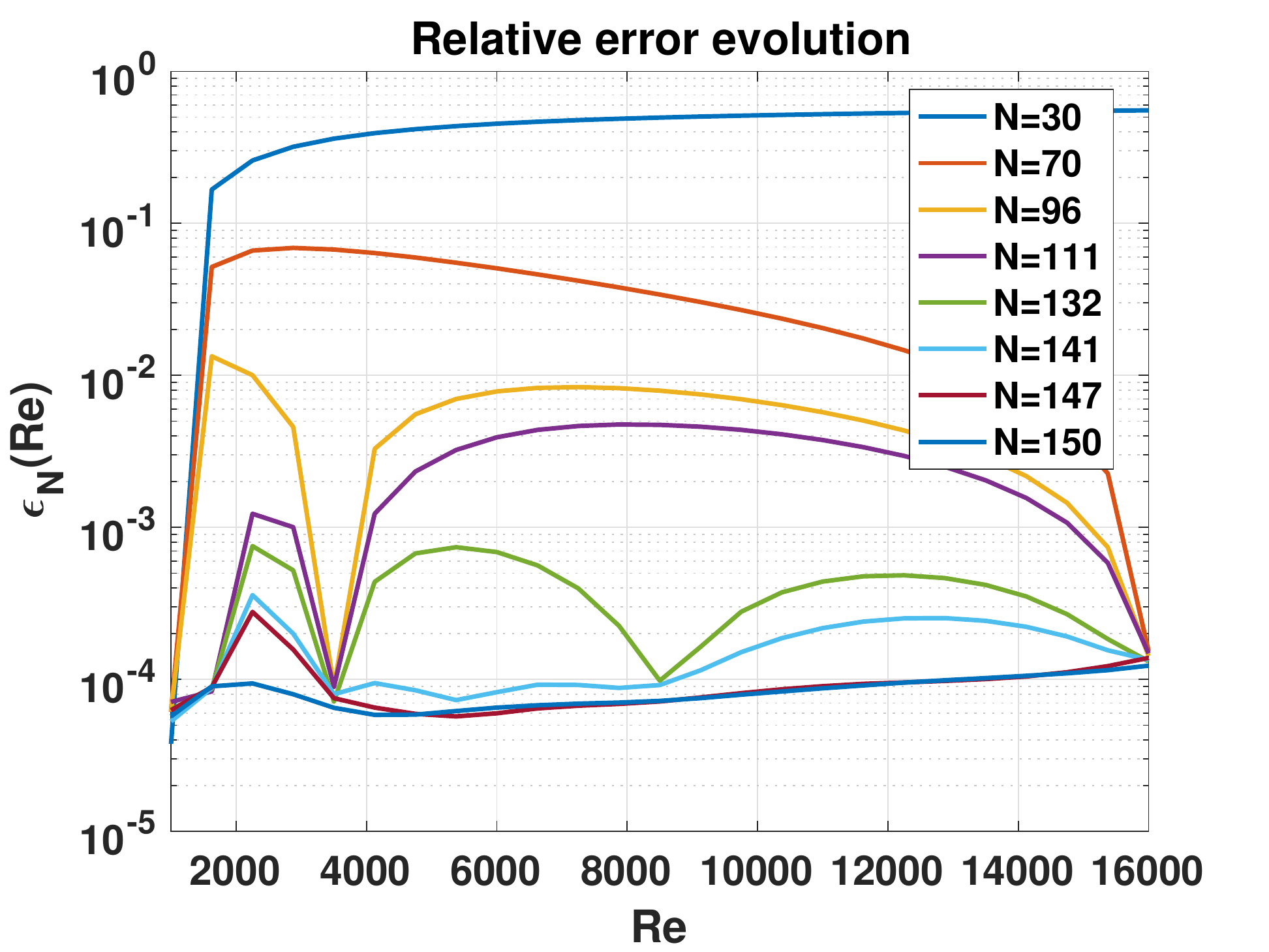}
    \caption{Error $\epsilon_N(\mu)$ in each iteration of the POD+Greedy algorithm, using the exact error as error indicator.}
    \label{fig:error 1POD}
\end{figure}

In Table \ref{tab:PODgreedy1}, we show the comparison of the errors between the reduced and the FOM solution, using the indicator $\Delta_N(\mu)$ introduced in \eqref{eq::Kolmogorov_apost_k2} (upper table) and the exact error $\epsilon_N(\mu)$ (lower table) for the selection of the new parameter value $\mu^*_N$ at each step. With some abuse of notation, but looking for a meaningful one, here we denote by $N$ the actual dimension of the combined reduced spaces $Y_N\times M_N$, instead of the iteration of the Greedy algorithm.  In Figures \ref{fig:errorEvolution 1POD} and \ref{fig:error 1POD}, we can see that the exact relative error and the number of RB basis are similar if we use either the indicator $\Delta_N(\mu)$ or the exact error $\epsilon_N(\mu)$.

\begin{figure}
 \centering
    \centering
    \includegraphics[width=0.8\textwidth]{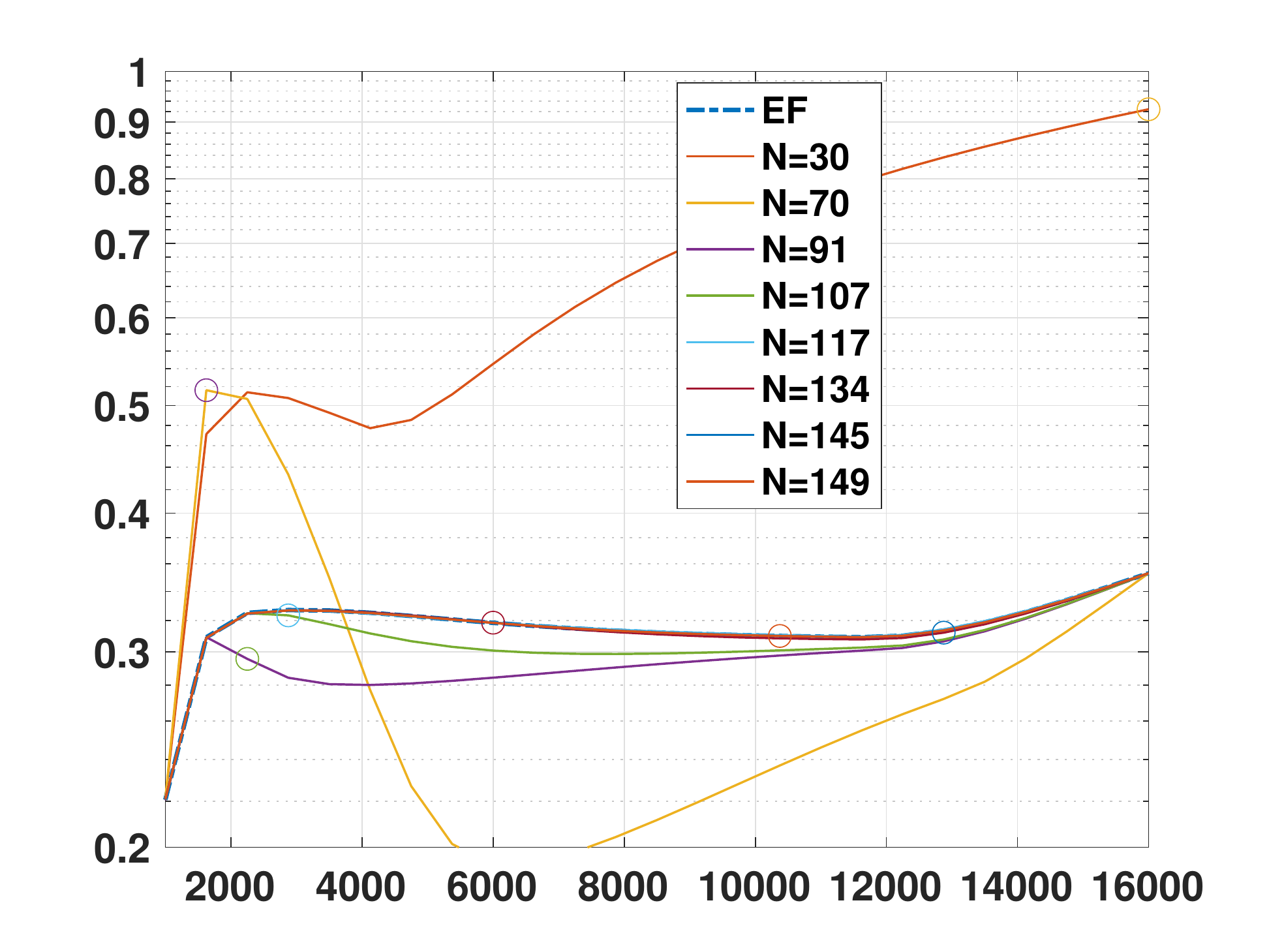}
    \caption{Evolution of  the ROM estimator, $\Delta_N(\mu)$, in each iteration of the Greedy algorithm, versus the FOM estimator, $\Delta_h(\mu)$. For each iteration, we plot the error indicator for every parameter in the range, and we highlight with a circle the parameter selected for the next iteration.}
    \label{fig:estimateEvolution 1POD}
\end{figure}

As previously said, since the reduced solution is built from the FE approximation, we should not expect that $\Delta_N(\mu)$ tends to $0$ as the dimension of the reduced space $N$ grows, it should rather converge to $\Delta_h$. We observe in Figure \ref{fig:estimateEvolution 1POD} that indeed $\Delta_N(\mu)$ approaches $\Delta_h$ as $N$ grows. In Figure \ref{fig:errorEvolution 1POD}, we show the relative error $\epsilon_N(\mu)$ for $\mu=\{1\,000,1\,625,\dots,16\,000\}$ at each POD+Greedy algorithm iteration. In the last iteration, the maximum relative error is nearly $10^{-4}$.

In Table \ref{tab:tablaresumen1POD}, we show, for various parameter values inside the parameter range different from the trial values, the computational time along with the speed-ups obtained, the values of the error indicators, and the error between the RB and FE solutions at the final time $T_f=30$. For the elaboration of this table, we have used the reduced basis model provided by the last iteration of the Algorithm \ref{alg:basic:1PODgreedySupremizer}, this is, the eight iteration, with $N=149$, shown on the upper table of Table \ref{tab:PODgreedy1}.

We can see that the \textit{a posteriori} error indicator of the reduced solution coincides at least up to the second decimal with the \textit{a posteriori} error indicator of the full-order solution for every tested parameter. Furthermore, We observe speeds-up of nearly 18. Speeds-up of one or two orders of magnitude are standard for ROM solution of parametric linear parabolic problems, here we extend them to LES turbulence models. As well, we obtain relative errors below of $10^{-4}$.

\begin{table}[ht]
\centering
\begin{tabular}{|c|ccccc|}
    \hline 
    $\mu$ & 3000 & 6000 & 9000 & 12000 & 15000 \\ \hline
    $T_{FE}$ & 59.11s & 59.05s & 58.80s & 58.82s & 59.48s \\
    $T_{RB}$ &  3.28s &  3.29s &  3.26s &  3.23s &  3.23s \\
    Speed-up & 18.02  & 17.95  & 18.04  & 18.21  & 18.41  \\ \hline
    $\Delta_N(\mu)$ & 0.327244 & 0.318597 & 0.311507 & 0.309537 & 0.337665 \\
    $\Delta_h(\mu)$ & 0.326930 & 0.318601 & 0.311539 & 0.309520 & 0.337642 \\
    $\epsilon_N(\mu)$ & $7.93\cdot10^{-5}$ & $6.05\cdot10^{-5}$ & $7.70\cdot10^{-5}$ & $8.37\cdot10^{-5}$ & $10.83\cdot10^{-5}$\\
    \hline
\end{tabular}
\caption{Validation of RB model.}
\label{tab:tablaresumen1POD}
\end{table}

\subsection{Equispaced sampling}

The objective of this second test is to compare the results obtained in the previous test with the ones obtained when considering equispaced parameter values, instead of considering a Greedy algorithm for selecting them.

\begin{figure}
    \centering
    \includegraphics[width=0.8\textwidth]{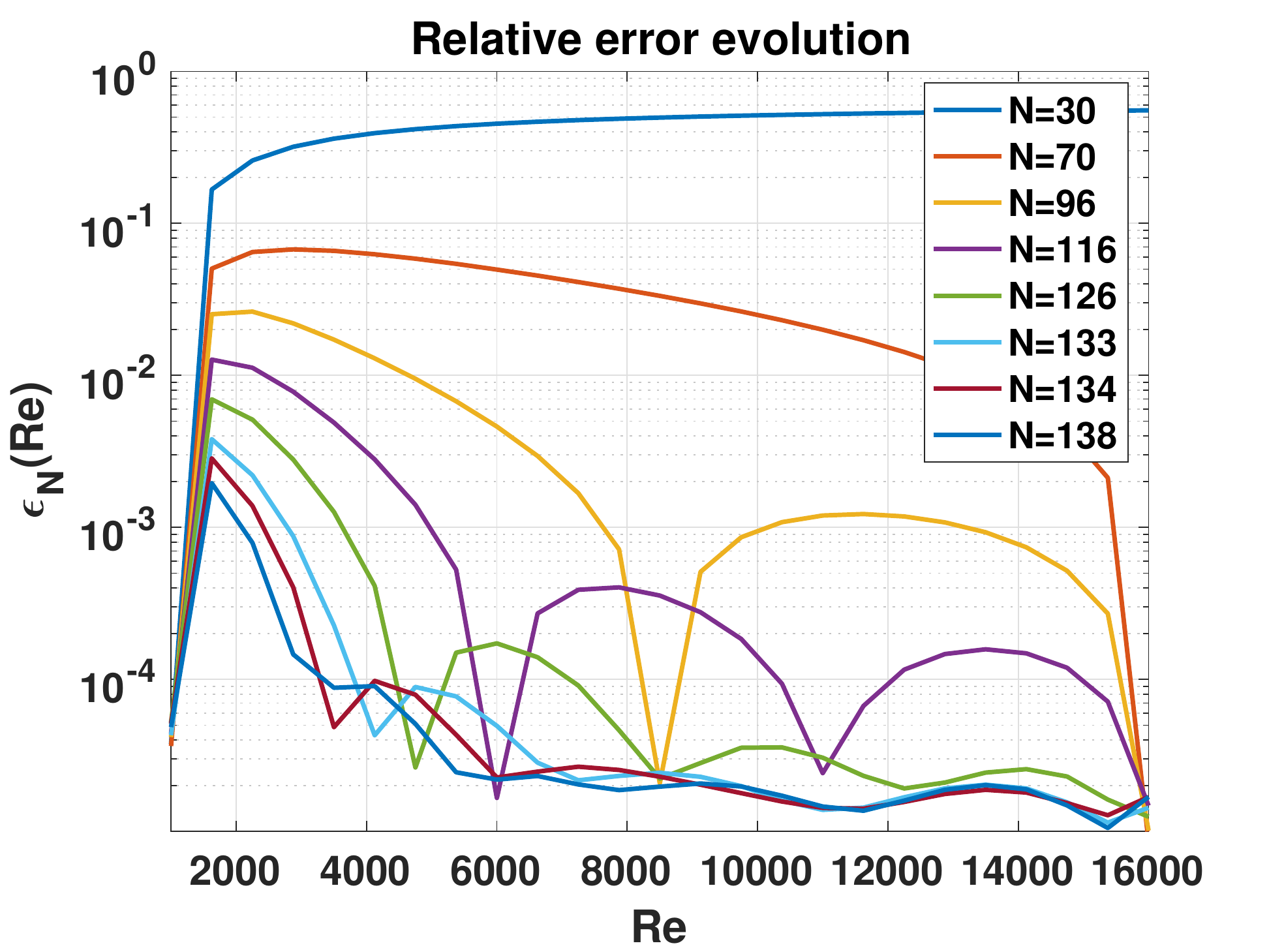}
    \caption{Convergence story of POD, using equispaced parameter values.}
    \label{fig:error equispaced}
\end{figure}

In Table \ref{tab:PODgreedyequispaced}, we present the maximum relative error and the $l_2$-norm of the relative error in the parameter range. We also present in Figure \ref{fig:error equispaced} the evolution of the relative error in the range of the parameter, we can see that the lagest errors appear in the lower Reynolds number range, at which the decrease is slower as $N$ increases.

\begin{table}[htb]
    \centering
    \begin{tabular}{|c|c|c c|}
        \hline
        It. & N & $\max_{Re}\epsilon_N(Re)$ & $l_2(\epsilon_N(Re))$ \\ \hline
        1   & 30  & $3,65\cdot 10^{-1}$ & $5,53 \cdot 10^{-1}$ \\ 
        2   & 70  & $6,74\cdot 10^{-2}$ & $1,95 \cdot 10^{-1}$ \\ 
        3   & 96  & $2,63\cdot 10^{-2}$ & $4,95 \cdot 10^{-2}$ \\ 
        4   & 116 & $1,27\cdot 10^{-2}$ & $1,96 \cdot 10^{-2}$ \\ 
        5   & 126 & $6,96\cdot 10^{-3}$ & $9,18 \cdot 10^{-3}$ \\ 
        6   & 133 & $3,81\cdot 10^{-3}$ & $4,49 \cdot 10^{-3}$ \\ 
        7   & 134 & $2,85\cdot 10^{-3}$ & $3,20 \cdot 10^{-3}$ \\ 
        8   & 138 & $1,96\cdot 10^{-3}$ & $2,12 \cdot 10^{-3}$ \\ \hline
    \end{tabular}
    \caption{Convergence of the POD algorithm, using equispaced parameter values.}
    \label{tab:PODgreedyequispaced}
\end{table}

\begin{figure}
 \centering
\begin{subfigure}{0.75\textwidth}
    \centering
    \includegraphics[width=\textwidth]{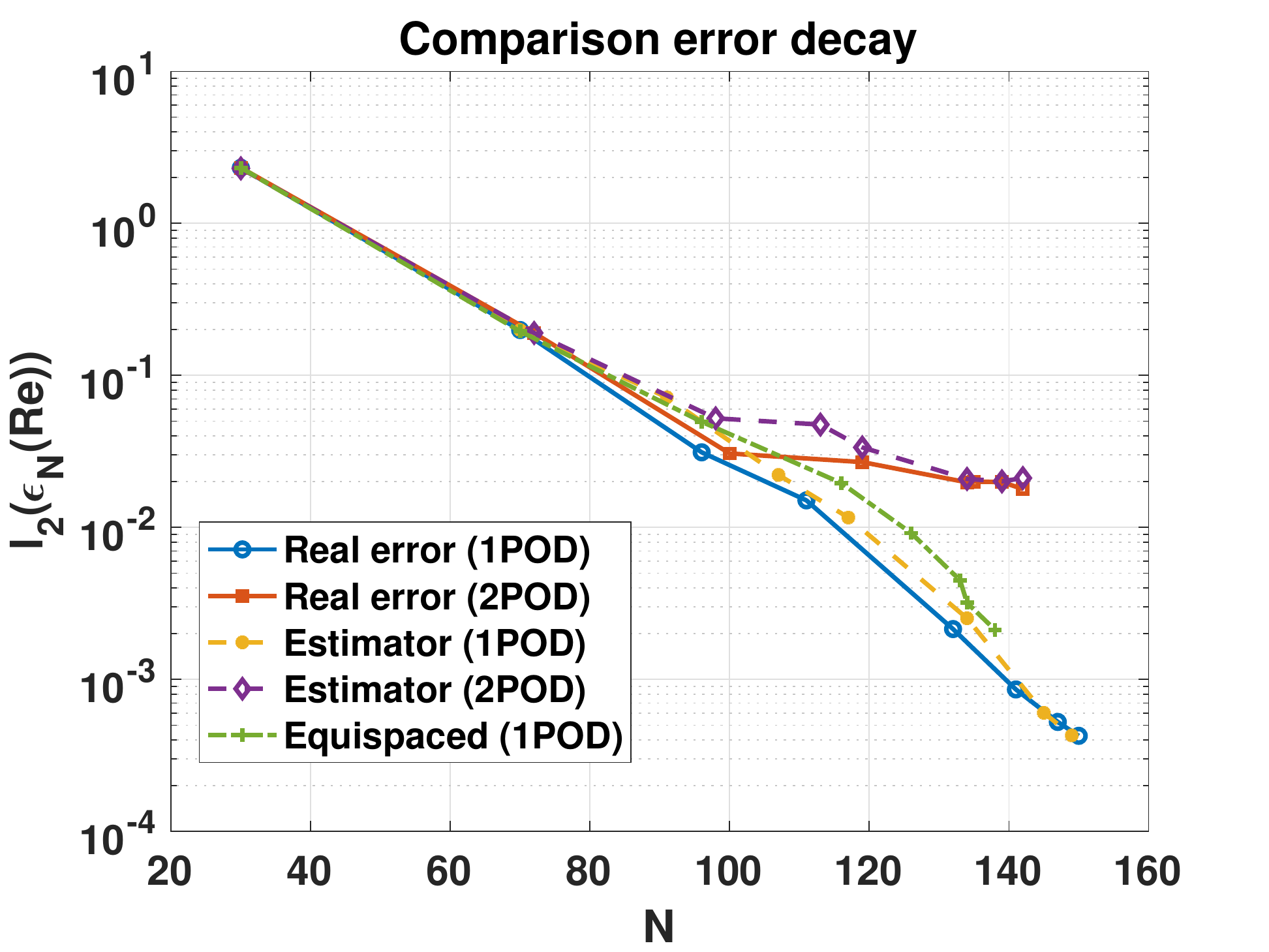}
    \caption{Comparison of relative $l_2$-norm decay.}
    \label{fig:Comparison decay l2}
\end{subfigure}
\begin{subfigure}{0.75\textwidth}
    \centering
    \includegraphics[width=\textwidth]{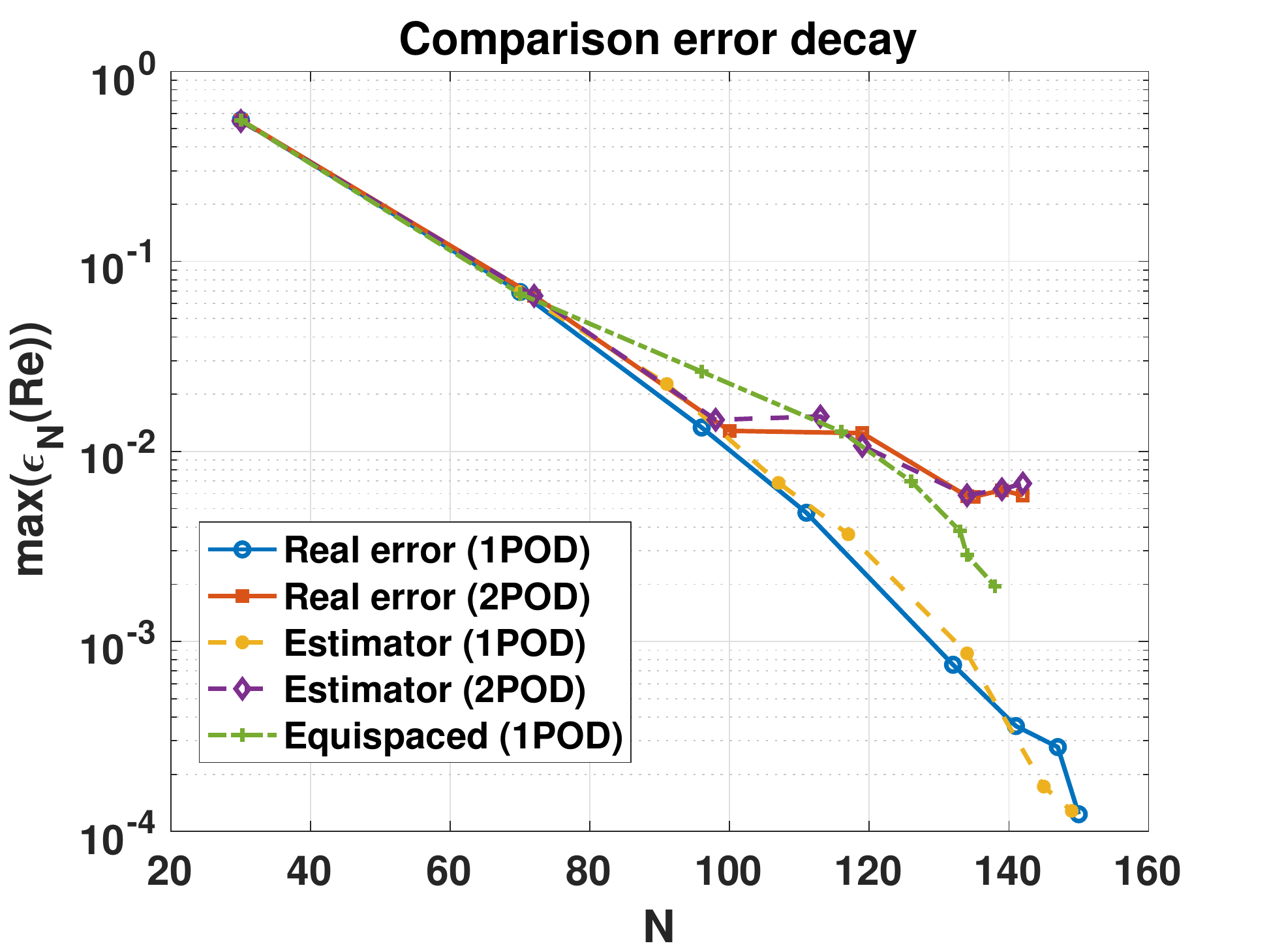}
    \caption{Comparison of relative $l_\infty$-norm decay.}
    \label{fig:Comparison decay max}
\end{subfigure}
\caption{Comparison of the relative error decay for the different methods. In both figures, the ROM errors obtained using the real error as erorr estimator appear in continuous line, in blue circles (1POD) and in red squares (2POD), the ones obtained by means of the a posteriori error indicator appear in dashed lines, in yellow bullets (1POD) and in purple diamonds (2POD), and the ones obtained by means of the equispaced parameters appear in green dash-dotted line with crosses (1POD).}
\label{fig:Comparisons}
\end{figure}

Finally, we compare all the presented procedures in Figure \ref{fig:Comparisons}. We present, in Figure \ref{fig:Comparison decay l2}, the comparison of the decay of the $l_2$-norm of the relative error over the parameter range and, in Figure \ref{fig:Comparison decay max} the comparison of the decay of the maximum relative error over the parameter range. We observe a spectral convergence with respect to the dimension of the reduced space $N$. We also observe that the \textit{a posteriori} error indicator based on Kolmogorov's law and the one-POD procedure presented in Algorithm \ref{alg:basic:1PODgreedySupremizer}, provides very similar results compared to the ones obtained using the exact error, in both $l_2$ and maximum relative error decay. Also, we observe that it provides better results than just selecting equispaced parameters in the reduced basis construction, specially in the maximum relative error.  We can also see that the procedure considering just one POD provides better results than the procedure in which two PODs were performed.


\section{Conclusions}\label{sec:Conclusions}

In this work, we have introduced an \textit{a posteriori} error indicator for the reduced-basis numerical solution of turbulent flows at statistical equilibrium. The indicator is based upon the Kolmogorov turbulence theory for turbulent flows, and is applicable to any numerical discretization and any LES turbulence model we are working with, as it only involves the physics of the turbulent flow. 

This allows to overcome the serious technical difficulties related to the construction of suitable specific \textit{a posteriori} error estimator for the Reduced Basis discretization for each problem, that should be specific to each actual kind of numerical discretisation.

We have validated this indicator with an academic numerical test considering the unsteady Smagorinsky turbulence model for 2D flows with well developed $k^{-5/3}$ spectrum, in which we have observed an exponential decay of the $l_2$ and $l_\infty$ relative errors with respect to the dimension of the reduced space. Also, the reached error values are quite close to those obtained when using of the exact error as an estimator for the RB construction. We have obtained an speed-up of the computations of nearly 18, what it is standard for parametric linear evolution problems, and we extend here to turbulent flows. We have also stated that considering just a one-POD procedure provides better results than considering a two-POD procedure for the construction of the reduced basis.

Therefore, this novel \textit{a posteriori} error indicator is a useful tool to the construction of the reduced models of turbulent flows, as it can be applied no matter what numerical discretization is used.

In future works, we intend to apply this novel error indicator to 3D flows and different LES models. We expect to provide a useful tool to dramatically speed-up the computation of turbulent flows of industrial interest.

\section*{Acknowledgements}
This work has been funded by the Spanish Government Project PID2021-123153OB-C21 and European Union’s Horizon 2020 research and innovation programme under the Marie Sklodowska–Curie Actions, grant agreement 872442 (ARIA).

\bibliography{Ref_RBM_new}
\bibliographystyle{siam}

\end{document}